\documentclass[10pt,preprintnumbers,twocolumn,amsmath,amssymb,nofootinbib,superscriptaddress]{revtex4-1}
\usepackage[latin1]{inputenc}
\usepackage{slashed}
\usepackage{textcomp}
\usepackage{amsfonts}
\usepackage{indentfirst}
\usepackage{color}
\usepackage{hyperref}
\usepackage{subcaption}
\usepackage[normalem]{ulem} %\sout

\newcommand{\be}{\begin{equation}}
\newcommand{\ee}{\end{equation}}
\newcommand{\bea}{\begin{eqnarray}}
\newcommand{\eea}{\end{eqnarray}}

\newcommand{\dd}{\mathrm{d}}

\usepackage{graphicx,graphics}
\graphicspath{{figures/}}

\begin{document}

\hfill{KCL-PH-TH/2024-49}

\title{Tunnelling and the Casimir effect on a $D$-dimensional sphere}

\author{Jean Alexandre} 
\author{Drew Backhouse}  

\affiliation{Theoretical Particle Physics and Cosmology, King's College London, WC2R 2LS, UK}

\begin{abstract}

Two fundamental signatures of Quantum Mechanics are tunnelling and the Casimir effect. We examine the ground state energetic properties of a scalar field confined on a $D$-dimensional sphere, and subjected to these two effects. We focus on $D=2$ and $D=3$, with a non-minimal coupling of a massless scalar field to curvature, which provides a radius-dependent effective mass. This scenario allows tunnelling to be more important than the Casimir effect, in a certain regime of parameters, and potential implications in Early Cosmology are discussed for the case $D=3$, which could avoid a cosmological singularity.

\end{abstract}

\maketitle

{\it Introduction}\\
The Casimir effect \cite{Casimir:1948dh} is the manifestation of the effects of boundary conditions on the vacuum fluctuations of a quantum field.
In its original observational configuration \cite{Casimir:1947kzi}, two parallel conducting plates are placed in the presence of the electromagnetic field and an attractive force between them is observed.
This force originates from a reduced vacuum energy between the boundaries due to the quantisation of vacuum fluctuations, an effect that also leads to the violation of the null energy condition (NEC - see \cite{Rubakov:2014jja} and \cite{Kontou:2020bta} for reviews).
Since its initial discovery, the Casimir effect has been calculated for a variety of different fields and boundaries, where it is found to be highly sensitive to their geometry and topology, in some cases even changing sign and thus satisfying the NEC (see \cite{Bordag} for a review).
Such an effect may even be present in the complete absence of boundaries, since the identification conditions of a compact topology play the same role as boundary conditions.
The latter effect is of interest within early universe cosmology since it may provide a source of NEC violation, required for features such as a cosmological bounce (see \cite{Brandenberger:2016vhg} for a review). We describe in the paragraph {\it Discussions} at the end of this letter the potential applications of this work within this field.

As detailed further below, another source of NEC violation due to the presence of boundaries/identification conditions has been under recent study and originates from a massive scalar field tunnelling between degenerate vacua \cite{Alexandre:2012ht,Alexandre:2022qxc,Alexandre:2023bih,Alexandre:2023iig,Alexandre:2023pkk,Alexandre:2024htk,Ai:2024taz}. Unlike the Casimir effect, tunnelling effects always lead to NEC violation, although they are exponentially suppressed by the spatial volume.
The Casimir effect however is at most exponentially suppressed by the length scale, and therefore dominates over tunnelling.

\begin{figure}
     \centering
     \begin{subfigure}{\linewidth}
         \centering
         \includegraphics[width=0.85\linewidth]{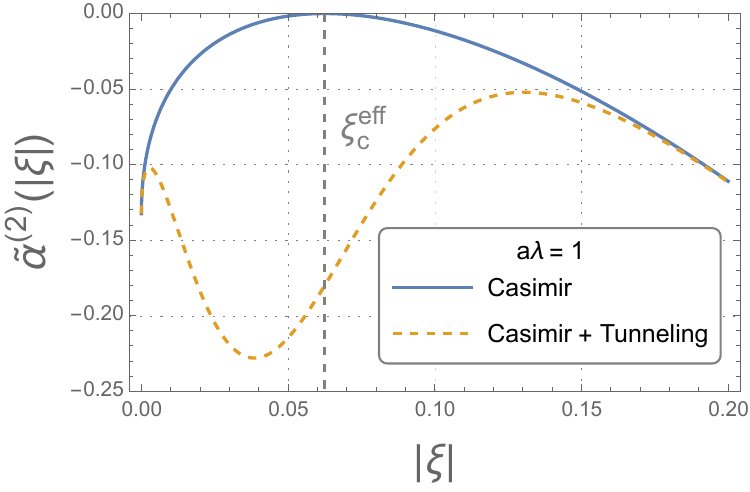}
         \caption{2-dimensions for $a\lambda=1$}
         \label{fig:S2}
     \end{subfigure}
     \vspace{0.1cm}\\
     \begin{subfigure}{\linewidth}
         \centering
         \includegraphics[width=0.85\linewidth]{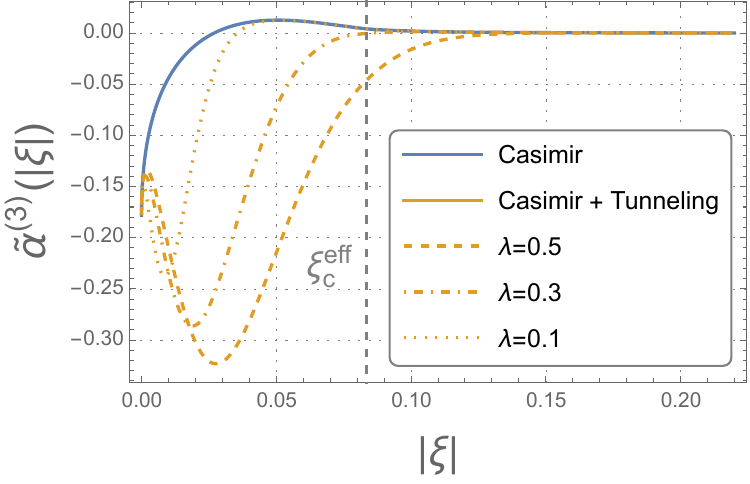}
         \caption{3-dimensions for $\lambda=0.3,0.2,0.1$.}
         \label{fig:S3}
     \end{subfigure}
        \caption{The coefficient of the vacuum energy $E_0^{(D)}=\Tilde{\alpha}^{(D)}(|\xi|)/a$ as a function of the non-minimal coupling $|\xi|$.}
        \label{fig:three graphs}
\end{figure}

In this letter, we instead focus on the effects of a massless scalar field tunnelling on a $D$-dimensional sphere, with an
effective mass generated by a general non-minimal scalar/curvature coupling $\xi$,
and find a regime where tunnelling effects are dominant. In particular, with the inclusion of tunnelling effects, we find the following two modifications to the parameter space where the NEC is violated on a D-dimensional sphere versus the Casimir effect alone:\\
$\bullet$ On a 2-dimensional sphere, NEC violation occurs for all values of $\xi$, and is far greater around the conformal effective value (c.f. Fig. \ref{fig:S2});\\
$\bullet$ On a 3-dimensional sphere, the value of the coupling $\xi$ for which NEC violation begins is higher and the violation is far greater (c.f. Fig. \ref{fig:S3}).\\

{\it Ground state from degenerate vacua}\\
Consider a symmetric double-well potential, with degenerate vacua and typical frequency $\omega$.
In Quantum Mechanics (QM), tunnelling between these vacua leads to a symmetric ground state wave function, 
and a vacuum energy which is lower than the vacuum energy $\omega/2$ in the individual wells. The difference in energy is proportional 
to $\omega\exp(-S_{inst}^0)$, where $S_{inst}^0$ is the action of the instanton configuration relating the two 
wells\footnote{The exponential dependence in $S_{inst}^0$ 
arises from the resummation over zero-modes of the instanton dilute gas, 
corresponding to the invariance of the total action describing the gas under translation of each individual instanton
(see \cite{Kleinert:2004ev} for a review based on the path integral approach).}.

In Quantum Field Theory (QFT), these features generalise to Euclidean-time-dependent instantons, 
but the corresponding action $S_{inst}$ is proportional to the volume $V$ of the space which confines the field.
For this reason, tunnelling between degenerate vacua in QFT is suppressed for an infinite volume, 
and Spontaneous Symmetry Breaking occurs instead
\footnote{This situation is different for $O(4)$-symmetric instantons \cite{Coleman:1977py,Callan:1977pt}, 
which have a volume-independent action.}. 
For a finite volume though, tunnelling is allowed and, similarly to what happens in QM,
symmetry is restored and the ground state energy is lower than the bare minima. 
This is consistent with convexity of the effective potential \cite{Symanzik:1969ek,Coleman:1974jh,Iliopoulos:1974ur,Haymaker:1983xk,Fujimoto:1982tc,Bender:1983nc,Hindmarsh:1985nc,Plascencia:2015pga,Millington:2019nkw,Alexandre:2012ht}, 
and the construction of a convex One-Particle-Irreducible effective potential from tunnelling is explicitly shown in \cite{Alexandre:2022qxc,Alexandre:2023iig}.

Taking tunnelling into account, the true ground state energy can then be obtained via a semi-classical approximation for the partition function $Z$ in the ground state, based on the dilute instanton gas approximation \cite{Kleinert:2004ev}
\be
Z\simeq\sum_{n=0}^\infty F_n~e^{-nS_{inst}}~,
\ee
where $n$ is the number of instantons and anti-instantons with action $S_{inst}$, and $F_n$ is the one-loop fluctuation factor. 
This approximation assumes that the $n$-instanton configurations are decoupled from each other, they do not overlap in field space, 
and is valid in the limit of zero temperature $(\beta\to\infty)$ and weak self-coupling $(\lambda\ll1)$. 

The vacuum energy is
\be
E_0=\lim_{\beta\to\infty}\left\{\frac{-1}{V\beta}\ln Z\right\}~,
\ee
where $\beta$ is the total Euclidean time. If we start from the double-well potential
\be
U(\phi)=\frac{\lambda}{24}(\phi^2-v^2)^2~,
\ee
we obtain \cite{Alexandre:2023iig,Alexandre:2024htk}
\be\label{E0}
E_0\simeq E_{Cas}-v\sqrt{\frac{\lambda}{\pi}~S_{inst}}~\exp(-S_{inst})~.
\ee
In the above expression, $E_{Cas}$ is the Casimir energy corresponding 
to a free field oscillating in one potential well with minimum $\phi=\pm v$ and represents the renormalised vacuum energy following the omission of the curvature-independent UV divergence, corresponding to the vacuum energy of flat space \cite{Bordag}.

The instanton which connects the bare vacua $\pm v$ is homogeneous and depends on Euclidean time only. It can be decomposed as:  \\
$\bullet$ ``static" parts which asymptotically go to $\pm v$;\\
$\bullet$ a Euclidean-time-dependent ``jump"  which connects the two static parts.\\
The expression (\ref{E0}) is obtained in the QM approximation, where the space-dependence of quantum fluctuations above the jump
part of the instanton are ignored. The static parts of the instanton are quantised with spacetime dependence though, 
involving discrete momenta, and leads to the Casimir energy $E_{Cas}$.

The full one-loop quantisation of the instanton is calculated in \cite{Ai:2024taz} in the situation of a three-torus.
This calculation is based on the resolvent method \cite{Baacke:2008zx} and the spectral decomposition of the Green's function corresponding to the fluctuation operator, which takes into account discrete momenta for fluctuations above the instanton.
The result shows that, for a weak self-coupling $\lambda\ll1$,
the QM approximation is very good, 
and is therefore what we use here.\\

{\it $D$-dimensional sphere}\\
We now consider a $D$-dimensional sphere as a confining space, with a non-minimally coupled massless scalar field.
The non-minimal coupling provides an effective mass which depends on the geometry. As a consequence and as shown below, 
the tunnelling contribution is not exponentially suppressed when the radius of the sphere increases.

We consider the non-minimal coupling $\xi R\phi^2/2$, together with a self-interaction term, such that the bare action is
\be\label{S}
S=\int d\tau\int d{\bf x}\sqrt{g}\left(\frac{1}{2}\partial^\mu\phi\partial_\mu\phi
+\frac{\lambda}{24}\left(\phi^2+6\frac{\xi R}{\lambda}\right)^2\right)~.
\ee
If we choose a negative coupling $\xi$, the latter action provides an effective double-well potential for the massless scalar field. 
This action involves a term proportional to $R^2$, whose significance is left for the discussion.
For time-dependent fields only, which is the case of the relevant saddle points in this problem, we have then
\be
S=V\int d\tau\left(\frac{1}{2}(\dot\phi)^2+\frac{\lambda}{24}\left(\phi^2-\frac{6|\xi|R}{\lambda}\right)^2\right)~.
\ee
A negative coupling $\xi$ therefore leads to fluctuations above the two bare minima,
with effective mass squared $2|\xi|R$, corresponding to the effective non-minimal coupling to gravity $\xi^{eff}=2|\xi|$.

The single-instanton configurations are solutions of the equation of movement, they minimise the action and therefore have a dominant contribution in the path integral. They are of the form
\be
\phi_s(\tau)=\pm \sqrt{6\frac{|\xi| R}{\lambda}}\tanh\left((\tau-\tau_0)\sqrt\frac{|\xi|R}{2}\right)~,
\ee
where the centre of the instanton $\tau_0$ is not fixed, and is at the origin of the translational invariance, 
or equivalently the zero mode of the fluctuation operator. The action for one instanton is then
\be \label{D-sphereInstAction}
S_{inst}=S[\phi_s]=\frac{2\sqrt2}{\lambda}(|\xi|R)^{3/2}~V~.
\ee

{\it Two-dimensional sphere}\\
We now consider a two-dimensional sphere as a confining space, for which the Casimir energy is \cite{Herdeiro:2007eb}
\be \label{ERen-2sphere}
E^{(2)}_{Cas}=
\frac{\alpha^{(2)}\big(|\xi|\big)}{a}~,
\ee
where
\bea \label{alpha2}
\alpha^{(2)}\big(|\xi|\big)=
\begin{cases}
    \displaystyle
    \mu^{3/2}
    \Bigg(
    -\frac{1}{3}
    +\int_0^1
    \frac{2t\sqrt{1-t^2}}
    {e^{2\pi\mu t}+1}
    \dd t
    \Bigg)~,
    & \mu>0\\
    \\
    0~, &\mu=0\\
    \\
    \displaystyle
    -|\mu|^{3/2}
    \int_0^1
    t\sqrt{1-t^2}
    \tan\left(\pi |\mu| t\right)
    \dd t~,
    & \mu<0
\end{cases}\nonumber
\eea
where $\mu\equiv 4|\xi|-1/4$.

For the radius $a$, the volume is $V=4\pi a^2$ and the curvature scalar is $R=2/a^2$, such that the action for one instanton is
\be \label{Sinst-2sphere}
S_{inst}^{(2)}=
\frac{32\pi|\xi|^{3/2}}
{\lambda a}~.
\ee
The vacuum energy \eqref{E0} is then
\be \label{E0S2}
E_0^{(2)}=\frac{\tilde{\alpha}^{(2)}\big(|\xi|,\lambda a\big)}{a}~,
\ee 
where
\bea\label{alphatilde2}
\tilde{\alpha}^{(2)}\big(|\xi|,\lambda a\big)
&\simeq&\alpha^{(2)}\big(|\xi|\big)\\
&&-\frac{16\sqrt{3}|\xi|^{5/4}}{\sqrt{\lambda a}}
\exp\left(-\frac{32\pi|\xi|^{3/2}}{\lambda a}\right)~.\nonumber
\eea
which is plotted in Fig. \ref{fig:S2}.

\begin{figure}
    \centering
    \includegraphics[width=0.85\linewidth]{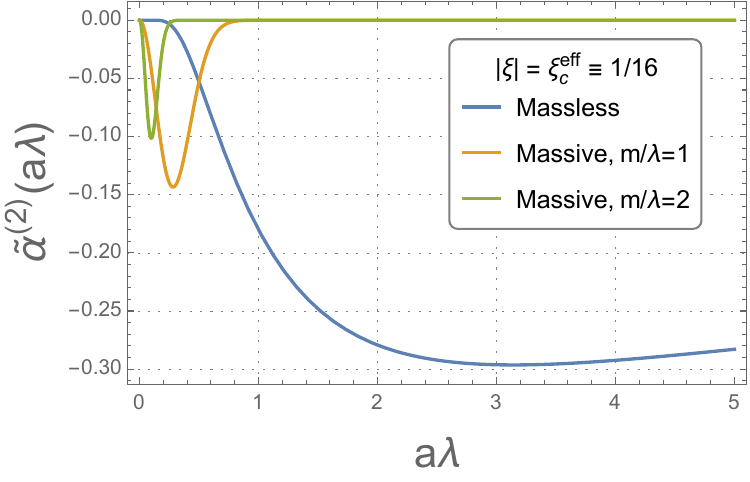}
    \caption{The coefficient of the vacuum energy \eqref{alphatilde2} in 2-dimensions for conformal effective coupling $|\xi|=1/16$ as a function of the dimensionless quantity $a\lambda$ and in comparison to the case of a minimally coupled massive scalar field \eqref{E0}. 
    In the situation of a massless field coupled non-minimally, tunnelling is not suppressed exponentially with the radius of the sphere.}
    \label{fig:alpha2alambda}
\end{figure}

As mentioned above, the tunnelling contribution is not exponentially suppressed when the radius of the two-sphere increases or for weak coupling,
unlike the case of a massive scalar field where the instanton action is proportional to $a^2m^3/\lambda$. 
A comparison between the massless and massive cases is displayed in Fig. \ref{fig:alpha2alambda}.
In addition, the Casimir energy vanishes for conformal effective coupling $\xi^{eff}=\xi_c=1/8$, such that $\xi=-\xi^{eff}/2=-1/16$. The tunnelling contribution remains however, such that the null energy condition is now violated for all values of $\xi$,
as shown in Fig. \ref{fig:S2}.\\

{\it Three-dimensional sphere}\\
We now consider a three-dimensional sphere as a confining space, for which the Casimir energy is \cite{Herdeiro:2007eb}
\be \label{ERen-3sphere}
E_{Cas}^{(3)}=
\frac{\alpha^{(3)}\big(|\xi|\big)}{a}~,
\ee
where
\be \label{alpha3}
\alpha^{(3)}\big(|\xi|\big)=
\begin{cases}
    \displaystyle
	\nu^2
    \int_1^\infty
    \frac{t^2\sqrt{t^2-1}}{e^{2\pi\nu t}-1}
    \dd t~,
    & \nu>0\\
    \\
    1/240 ~,&\nu=0\\
    \\
    \displaystyle
    \nu^2
    \int_0^\infty
    \frac{t^2\sqrt{t^2+1}}{e^{2\pi\nu t}-1}\dd t
    &\\
    \displaystyle
    \quad+\frac{\nu}{2}\int_0^1
    t^2\sqrt{1-t^2}\cot\Big(\pi\nu t\Big)\dd t~,
    & \nu<0
\end{cases}\nonumber
\ee
where $\nu\equiv 12|\xi|-1$.

For the radius $a$, the volume is $V=2\pi^2 a^3$ and the curvature scalar is $R=6/a^2$, such that the action for one instanton is
\be \label{Sinst-3sphere}
S_{inst}^{(3)}=
\frac{48\pi^2\sqrt{3}|\xi|^{3/2}}{\lambda}~.
\ee
The vacuum energy \eqref{E0} is then
\be\label{E0S3}
E_0^{(3)}=
\frac{\Tilde{\alpha}^{(3)}(|\xi|,\lambda)}{a}~,
\ee
where
\bea\label{alphaTotal3}
&&\Tilde{\alpha}^{(3)}(|\xi|,\lambda)\simeq
\alpha^{(3)}(|\xi|)\\
&&-\frac{24\sqrt{2\pi}\times3^{3/4}|\xi|^{5/4}}{\sqrt{\lambda}}
\exp\left(-\frac{48\pi^2\sqrt{3}|\xi|^{3/2}}{\lambda}\right)~,\nonumber
\eea
and is plotted in Fig. \ref{fig:S3}.

As in the situation of the two-sphere, the tunnelling contribution is not exponentially suppressed when the radius increases. In fact, the instanton action is independent of the length scale, such that the total vacuum energy has a simple $1/a$ dependence, like the Casimir energy. 
The tunnelling contribution is however suppressed for a weak self-coupling $\lambda$, but this may be counteracted with a sufficiently small value for the coupling $\xi$.\\

{\it Discussions}\\
In this work we consider the inclusion of tunnelling on a static $D$-dimensional sphere, in addition to the well-known Casimir effect. 
Tunnelling is achieved by considering a negative coupling $\xi$ in the bare theory, as well as a self-interaction for the scalar field, such that the bare theory exhibits two degenerate vacua, each with an effective curvature coupling of $\xi^{eff}=|\xi|/2$. 
Similarly to the Casimir effect, tunnelling between degenerate vacua contributes to a non-trivial ground state energy, but tunnelling always leads to NEC violation, independently of the parameters. In addition, the scenario studied here allows tunnelling to dominate over the Casimir effect, modifying the parameter space where the NEC is violated on a D-dimensional sphere, as exhibited in Figs. \ref{fig:S2} and \ref{fig:S3}.

Renormalising the ground state energy is achieved by the replacement of the bare vacuum energy with the Casimir energy. We note that the divergent contribution only coincides with the vacuum energy of flat space in the large $a$ limit, which is the procedure used in \cite{Herdeiro:2005zj}. However, the result obtained is identical to that via the more rigorous method of renormalising the gravitational sector, as was taken in later work by the same authors \cite{Herdeiro:2007eb}. 
Our work relies on the introduction of an $R^2$ term though, which modifies Einstein's equations in the situation where one considers dynamical gravity. This term leads to the well-known Starobinsky scenario \cite{Starobinsky:1980te}, and would require a more thorough analysis of renormalisation for the ground state energy.

A natural extension of the three-sphere example is to study Early Cosmology, where 
NEC violation could dynamically induce a cosmological bounce, in a scenario where the scale factor decreases. 
The Casimir energy of a conformally coupled massless scalar field in the Einstein static universe (ESU)
with spatial topology of a 3-sphere was considered in \cite{Ford:1975su,Ford:1976fn}, without NEC violation though, 
and the situation of a dynamical spacetime metric was studied in \cite{Mamaev:1980nj, Grib:1980pk,Zeldovich:1984vk}. 
The case of an ESU was then extended in \cite{Herdeiro:2005zj,Herdeiro:2007eb}, with a non-conformally coupled scalar field, 
and a repulsive Casimir force was found, violating the NEC and potentially sustaining a cosmological bounce. 
Similar studies involving tunnelling have been done in \cite{Alexandre:2023pkk,Alexandre:2023bih,Ai:2024taz}, 
assuming a three-torus as a confining space.
In these works, NEC violation indeed leads to a cosmic bounce, which is induced dynamically when the comoving volume decreases.
In the present study though, because of the $R^2$ contribution, we would need to study the possibility of a bounce with the modified Friedman-Lemaitre-Robertson-Walker equations, and these studies are left for future work.\\

\section*{Acknowledgments}
The Authors would like to thank Matthias Carosi, Eleni Kontou, Silvia Pla and Ben A. Stefanek for useful comments,
and the QFTCS Workshop III for facilitating discussions.
This work is supported by the Science and Technology Facilities Council (grant No.STFC-ST/X000753/1), and JA is also supported by 
the Leverhulme Trust (grant No. RPG-2021-299) as well as the Engineering and Physical Sciences Research Council (grant No. EP/V002821/1).
For the purpose of Open Access, the authors have applied a CC BY public copyright licence to any Author Accepted Manuscript version arising from this submission.

\bibliography{Bib}

\end{document}